# Mechanical properties of cubic boron nitride and diamond at dynamical pressure and temperature

S. M. Kastuar,[1] Z. -L. Liu,[2] S. Najmaei,[3] and C. E. Ekuma[1, a)]
[1)] *Department of Physics, Lehigh University, Bethlehem, Pennsylvania 18015, USA*
[2)] *College of Physics & Electric Information, Luoyang Normal University, Luoyang 471934, China*
[3)] *Sensors & Electron Devices Directorate, United States Army Research Laboratory, Adelphi, Maryland 20783, USA*

(Dated: 10 November 2023)

We report the mechanical properties of cubic boron nitride (c-BN) and diamond under the combined impact of dynamical pressure and temperature, calculated using *ab-initio* molecular dynamics. Our study revealed a pronounced sensitivity of the mechanical properties of c-BN to applied pressure. Notably, c-BN undergoes a brittle-to-ductile transition at ∼220 GPa, consistent across various dynamical temperatures, while diamond exhibits no such transition. Furthermore, the Vickers hardness profile for c-BN closely mirrors that of diamond across a spectrum of temperature-pressure conditions, highlighting c-BN's significant mechanical robustness. These results underscore the superior resilience and adaptability of c-BN compared to diamond, suggesting its potential as an ideal candidate for applications in extreme environments.

The last decade has seen a renaissance in the study of the mechanical properties of superhard materials and nanostructures. This resurgence can be attributed to two pivotal developments: advancements in predictive materials modeling[1,2] and innovations in experimental measurements[3,4]. Together, these methodologies offer unparalleled insights into the stability and bonding mechanisms that support the resistance of superhard systems to both elastic and plastic deformations. Such profound understanding has catalyzed the discovery of a plethora of new superhard materials[1,5–8]. Although the mechanical attributes of numerous materials have been meticulously examined using *ab initio* methodologies, the majority of these studies are mainly at zero temperature. Consequently, the interplay of dynamical pressure with temperature remains an area requiring further investigation. Typically, when characterizing the mechanical behavior of superhard materials, emphasis is placed on metrics such as shear or tension strain. Detailed studies of the dynamic effects of temperature and pressure on bulk properties, such as the Young's modulus and bulk modulus, are limited. This gap in research can be attributed to a lack of advanced computational techniques capable of accurately simulating the pressure-temperature phase diagram and the inherent challenges of conducting precise measurements at minuscule volume deformations. To harness the potential of superhard materials, especially in extreme environments like space technology and structural composite applications, it is essential to understand their temperature-dependent behavior. As materials' applications become more advanced and intricate, optimizing both functional and mechanical properties becomes critical. Achieving this optimization requires a comprehensive understanding of the effects of temperature and pressure on these materials' mechanical properties.

In this Letter, we investigate the structural stability and mechanical properties of cubic boron nitride (c-BN) and diamond under extreme conditions - namely high dynamical temperature and pressure - using *ab initio* molecular dynamics (AIMD). c-BN is isoelectronic to diamond and shares many of the diamond's exceptional chemical and physical properties (Figure 1(a)); it forms other stable analog polymorphic structures, e.g., h-BN, amorphous BN, and BN nanotubes to carbon-based structures.[9–11] However, due to the partial ionicity of the B-N bonds, c-BN possesses unique chemical and physical properties not attainable in diamond, such as suitability as a $p - n$ junction diode,[12] high stability in diverse extreme conditions such as oxidizing environments and in contact with Group VIII metals (Fe, Co, and Ni).[13–16] The outstanding combination of these chemical and physical properties with an ultra-wide bandgap > 5.0 eV that is transparent to the ultraviolet regime of the electromagnetic spectrum, underpins our current study. Our findings suggest that c-BN outperforms diamond, with an anticipated brittle-to-ductile transition under increased pressure and temperature.

The dynamical pressure and temperature-dependent mechanical properties were obtained using the thermal stress-strain data from *ab initio* molecular dynamics[17] with the ElasTool toolkit[18,19] using VASP[20] as the calculator. AIMD provides a reliable description of the time evolution of systems and often reveals non-intuitive temperature-dependent system configurations. We employed a 3×3×3 supercell with a 2×2×2 Monkhorst-Pack $k$-point grid for the reciprocal space sampling. The structures were initially optimized with the Perdew, Burke, and Ernzerhof (PBE) exchange-correlation functional[21] and then equilibrated in an isothermal isobaric (NPT) ensemble using the Langevin thermostat to maintain the temperature. The structures were further simulated un-

---

a)che218@lehigh.edu

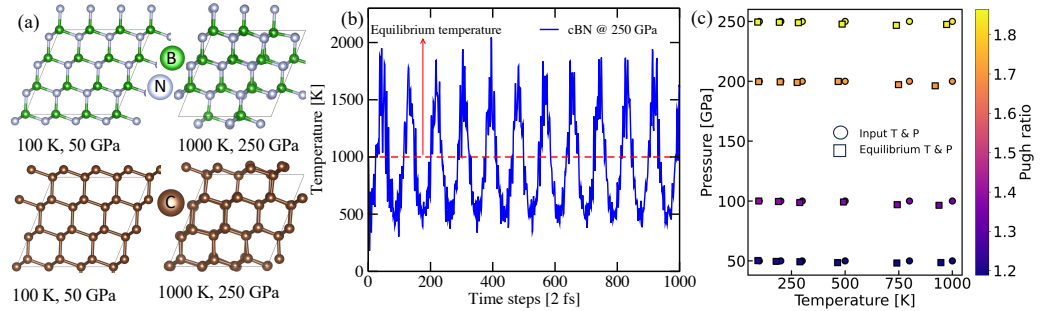

FIG. 1. (a) A representative crystal structure of c-BN and diamond at both low and high temperatures and pressures. (b) The dynamical steps at 250 GPa and 1000 K for c-BN. (c) Temperature-pressure profile for the Pugh ratio of c-BN. Circular and square markers, respectively, represent the input and equilibrium temperatures and pressures.

TABLE I. Mechanical properties of c-BN and diamond at zero temperature and various dynamical pressures. [a]

| Material | B | G | E | $\nu$ | $V_m$ | $\Theta_D$ |
|---|---|---|---|---|---|---|
| | | | 0 GPa | | | |
| **c-BN** | 376.38 | 380.31 | 853.47 | 0.12 | 11.48 | 1496.48 |
| **Diamond** | 436.12 | 516.03 | 1110.21 | 0.08 | 13.24 | 1751.61 |
| | | | 100 GPa | | | |
| **c-BN** | 711.48 | 510.37 | 1235.65 | 0.21 | 12.20 | 1694.69 |
| **Diamond** | 776.61 | 680.07 | 1579.24 | 0.16 | 14.07 | 1969.99 |
| | | | 250 GPa | | | |
| **c-BN** | 1154.73 | 618.83 | 1575.12 | 0.27 | 12.47 | 1828.63 |
| **Diamond** | 1221.63 | 832.06 | 2034.31 | 0.22 | 14.51 | 2137.91 |

[a] The bulk B and shear G modulus (in units of GPa) are obtained from the Voigt-Reuss-Hill approximations, E is Young's modulus in GPa, $\nu$ is the Poisson ratio, $V_m$ is the average sound velocity, and $\Theta_D$ is the Debye temperature.

der a canonical (NVT) ensemble using the Nose-Hoover thermostat to maintain the temperature. All calculations used a cutoff energy of 550 eV, a time step of 2 fs, and 1000 MD steps with the last 500 MD steps used to average the thermal stresses (see Figure 1b for the MD steps).

Our analysis begins with Table I, which displays mechanical properties computed at zero temperature across different dynamic pressures. The dynamic behavior is depicted in Figure 1c, where both temperature and pressure are equilibrated at every MD step. We note that under ambient conditions, our calculated data are in good agreement with previous experiments.[22,23] Our data indicates a systematic enhancement in the mechanical properties of the materials with increasing pressure. This observation suggests that as the material is subjected to compression, its atoms or molecules become more densely packed, leading to a minimized structural porosity and an increase in its overall resilience. Such behavior often hints at potential structural phase changes, similar to how carbon transforms into a harder diamond phase under specific high-pressure conditions. This amplified strength under pressure highlights not only the potential suitability for high-pressure applications, such as deep-sea exploration or aerospace, but also its increased resistance to external forces and deformations. We note that both structures show remarkable resilience within the parameter space explored (Figure 1a). However, while these initial findings are promising, it is paramount to further explore and understand the behavior under extreme compressive conditions, as there may be thresholds beyond which the material undergoes irreversible changes or potential structural collapse. Nevertheless, a nuanced understanding of pressure-dependent behavior is crucial to optimize its application in future technological advances. Under ambient pressure and dynamic temperature conditions, there was a slight decrease in mechanical properties (Table II). Intriguingly, the mechanical properties, including bulk, shear, and Young's modulus, are more sensitive to pressure, nearly tripling with a 250 GPa increase, than to temperature variations. In contrast, temperature up to 1000 K results in only a slight decrease in these mechanical values. The Zener ratio $A = 2C_{44}/(C_{11} - C_{12})$ measures elastic anisotropy; $A$ notably changes with pressure but only marginally with temperature. Specifically, for c-BN, $A$ increased from 1.70 to 2.00, and for diamond, from 1.30 to 1.60, as the pressure is increased from 50 to 200 GPa. In contrast, a temperature rise to 1000 K yields a minimal 2-3% reduction in $A$ for both, highlighting pressure's dominant role in their anisotropy[24].

We present in Figure 2, the temperature-pressure phase diagram for c-BN and diamond, characterized by Pugh's modulus ratio $\vartheta_r = B/G$, a widely recognized parameter to distinguish between brittle and ductile behavior in materials. A $\vartheta_r$ greater than 1.75 is generally indicative of ductile behavior, while a lower value suggests brittleness[25]. This detailed phase diagram re-veals the intrinsic mechanical response of the materials under dynamic conditions, effectively depicting the interplay between temperature and pressure. At zero temperature (see Figure 3), an increase in $\vartheta_r$ with pressure is observed for both materials. Specifically, we predict a brittle-to-ductile transition in c-BN at $\sim$220 GPa. The Poisson ratio can also be used as an indicator of a material's brittleness or ductility, with a critical value around 0.26 demarcating the two. Materials with a Poisson ratio smaller than 0.26 are typically brittle, whereas those with larger values are ductile.[26] In the case of c-BN, as pressure increases from 200 GPa to 250 GPa, $\nu$ surpasses this critical value (Table I), indicating a transition from brittle to ductile behavior, which is corroborated by the Pugh's modulus ratio. A purely covalent material typically exhibits a $\nu$ close to 0.10, underscoring the strong directional bonds. During deformation, the B-N bonds in c-BN elongate more slowly than the C-C bonds (Figure 1a), aligning with the observed decrease in reactivity as the material approaches the cubic-to-graphitic transition while retaining its sp$^3$ bonding up to the critical strain.[27] Extremely high temperatures (>1000 K) and high pressures (>300 GPa) pose computational and experimental challenges, but extrapolation could allow us to gain some insight into the phase space of temperature-pressure. The extrapolated data, characterized by an accuracy $R^2 \approx 0.98$, indicates a maximum ductility at $\sim$570 GPa for both the c-BN and diamond systems. It's noteworthy that pressures of such magnitude are often beyond the reach of most experimental setups and present formidable computational challenges due to profound crystal structure deformations. Notwithstanding, a critical observation is that diamond consistently retains its brittleness across the entire temperature-pressure phase diagram. The brittle-to-ductile transition of c-BN highlights its potential superiority over diamond, particularly under extreme conditions that demand high resilience and mechanical stability. In environments where both pressure tolerance and ductility are crucial, c-BN could offer advantages over diamond for certain applications, underscoring its potential for next-generation high-demand environments and advanced material applications.

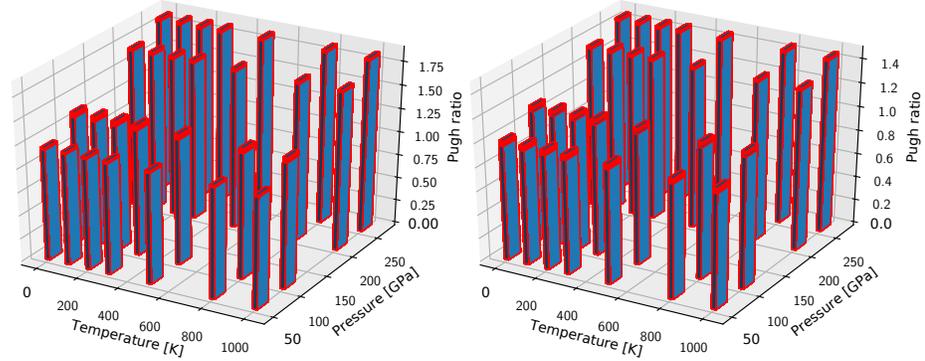

FIG. 2. Temperature-pressure phase diagram illustrated by the Pugh ratio $\xi_r$ (B/G): c-BN (left panel) and diamond (right panel). Diamond consistently exhibits brittleness across elevated pressure and temperature, whereas c-BN is predicted to undergo a brittle to ductile transition, seemingly unaffected by temperature variations.

TABLE II. Mechanical properties of c-BN and diamond at zero pressure and various dynamical temperatures. [a]

| Material | B | G | E | $\nu$ | $V_m$ | $\Theta_D$ |
|---|---|---|---|---|---|---|
| temperature at 300 K | | | | | | |
| c-BN | 345.39 | 358.09 | 798.36 | 0.12 | 11.27 | 1456.85 |
| Diamond | 417.00 | 497.13 | 1067.28 | 0.07 | 13.09 | 1722.93 |
| temperature at 500 K | | | | | | |
| c-BN | 328.39 | 347.94 | 771.39 | 0.11 | 11.15 | 1437.62 |
| Diamond | 395.55 | 482.85 | 1029.60 | 0.07 | 12.99 | 1701.08 |
| temperature at 1000 K | | | | | | |
| c-BN | 304.84 | 326.43 | 721.69 | 0.11 | 10.91 | 1396.62 |
| Diamond | 369.19 | 458.56 | 972.89 | 0.06 | 12.77 | 1662.36 |

[a] Parameters and definitions are the same as in Table I.

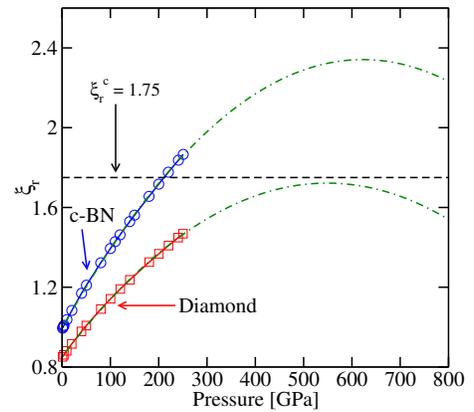

FIG. 3. Pugh's modulus ratio $\xi_r$ (B/G) for c-BN and diamond at zero temperature and various dynamical pressures. Notably, a brittle-to-ductile transition is predicted for c-BN around 220 GPa, while diamond consistently exhibits brittle behavior. The dashed green curve delineates a polynomial fit to our computational data with an accuracy of $\sim$0.98 for both materials. The parameter $\xi_r^c$ serves as the criticality between brittleness and ductility.

To further quantify the mechanical properties of the materials under dynamic temperature-pressure conditions, we computed the Vickers hardness, as illustrated in Figure 4. The hardness of both c-BN and diamond-like systems is effectively described by the relationship between the Vickers hardness ($H_\nu$), G, and B: $H_\nu = 2(G/\vartheta_r^2)^{0.585} - 3$[28]. The experimental and computational determinations of Vickers hardness of materials exhibit a considerable range of values, influenced by the specific methods used. Variability arises from the difficulties in precisely measuring indentation dimensions, fractures occurring within the crystal samples, the properties of the indenter tip, and the creation of indistinct indentations. For example, the reported $H_\nu$ values for diamond vary between 60 and 200 GPa[28–30]. Our empirical equation yields an $H_\nu$ of ∼90 GPa at zero pressure, aligning with the range of these previously reported experimental and computational results. The temperature-dependent $H_\nu$ at various dynamical pressures indicates that both c-BN and diamond consistently show a reduction in ($H_\nu$) as the pressure increases, with only slight variations due to temperature changes. Furthermore, with increased pressure, the $H_\nu$ of c-BN closely mirrors the entire $H_\nu$ pressure profile observed for diamond, highlighting the potential of c-BN to match the hardness levels exhibited by diamond.

In summary, we employed AIMD to probe the mechanical behaviors of c-BN and diamond under dynamical high temperature and pressure conditions. Our data reveal a pronounced sensitivity in the mechanical properties of both materials to external pressure. A notable divergence between the two materials was observed: cBN manifested a brittle-to-ductile transition at 220 GPa, a phenomenon that remained consistent irrespective of temperature fluctuations, distinguishing it from diamond, which lacked this adaptability. Moreover, the temperature-pressure Vickers hardness trajectory of c-BN mirrors that of diamond, highlighting its comparably robust nature. These findings collectively suggest that c-BN, with its superior mechanical resilience compared to diamond, is a promising candidate for applications requiring durability under extreme pressure and temperature conditions. This study paves the way for harnessing the potential of c-BN in next-generation high-performance applications in extreme environments. This finding is also relevant for the current research in geologic minerals under pressure at often high temperatures.

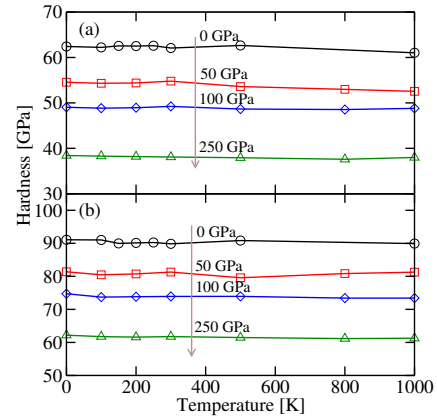

FIG. 4. Temperature-dependent Vickers hardness $H_\nu$ at various dynamical pressures for (a) c-BN and (b) diamond obtained with the relation $H_\nu = 2(G/\vartheta_r^2)^{0.585} - 3$, where $\vartheta_r = B/G$ is the Pugh's modulus ratio. The arrow shows the systematic decrease of $H_\nu$ as pressure is increased.

## ACKNOWLEDGMENT

This research is supported in part by the National Science Foundation DMR-2202101 and the National Natural Science Foundation of China (11974091). S.M.K acknowledges the Lee Graduate Fellowship from the College of Arts and Sciences, Lehigh University. Computational resources were provided by the DOD HPCMP at the Army Engineering Research and Development Center in Vicksburg, MS.

## AUTHOR DECLARATIONS

### Conflict of Interest

The authors have no conflicts to disclose.